\documentclass[english,prl, reprint, superscriptaddress, showkeys, showpacs, footinbib, raggedbottom, balancelastpage]{revtex4-1}
\usepackage[T1]{fontenc}
\usepackage[latin9]{inputenc}
\usepackage{booktabs}
\usepackage{amsmath}
\usepackage{amsthm}
\usepackage{graphicx}

\makeatletter

\providecommand{\tabularnewline}{\\}

\theoremstyle{plain}
\newtheorem*{prop*}{\protect\propositionname}
\theoremstyle{plain}
\newtheorem*{cor*}{\protect\corollaryname}

\usepackage{ytableau}
\definecolor{blueviolet}{rgb}{0.2, 0.2, 0.6}
\usepackage[pdftex,          
    bookmarks=false,          
    colorlinks=true,
    allcolors=blueviolet,
    pdfstartview={FitH},  
    ]{hyperref}
\usepackage{txfonts}
\usepackage[preset=reset]{phfnote}

\makeatother

\usepackage{babel}
\providecommand{\corollaryname}{Corollary}
\providecommand{\propositionname}{Proposition}

\begin{document}
\phfnoteSaveDefs{origcmds}{H,c,k,l,b,d,r,aa,u,v,t,o}

\global\long\def\bra{\langle}
\global\long\def\ket{\rangle}
\global\long\def\half{{\textstyle \frac{1}{2}}}
\global\long\def\third{\frac{1}{3}}
\global\long\def\dx{\frac{\partial}{\partial x}}
\global\long\def\dxd{\frac{\partial}{\partial\dot{x}}}
\global\long\def\thrha{\frac{3}{2}}
\global\long\def\sq{\sqrt{2}}
\global\long\def\sqinv{\frac{1}{\sqrt{2}}}
\global\long\def\up{\uparrow}
\global\long\def\do{\downarrow}
\global\long\def\p{\partial}
\global\long\def\dqi{\frac{\partial}{\partial q_{i}}}
\global\long\def\dqid{\frac{\partial}{\partial\dot{q}_{i}}}
\global\long\def\a{\alpha}
\global\long\def\b{\beta}
\global\long\def\g{\gamma}
\global\long\def\c{\chi}
\global\long\def\d{\delta}
\global\long\def\m{\mu}
\global\long\def\n{\nu}
\global\long\def\z{\zeta}
\global\long\def\l{\lambda}
\global\long\def\e{\epsilon}
\global\long\def\x{\chi}
\global\long\def\r{\rho}
\global\long\def\t{\theta}
\global\long\def\G{\Gamma}
\global\long\def\D{\mathcal{D}}
\global\long\def\O{\mathcal{O}}
\global\long\def\L{\mathcal{L}}
\global\long\def\T{\mathcal{T}}
\global\long\def\I{{\cal I}}
\global\long\def\dg{\dagger}
\global\long\def\k{\kappa}
\global\long\def\P{\mathcal{P}_{\!\!\!{\scriptscriptstyle \infty}}}
\global\long\def\R{{\cal R}}

\global\long\def\LE{{\cal L}_{\textnormal{\textsf{eff}}}}
\global\long\def\HE{H_{\textnormal{\textsf{eff}}}}
\global\long\def\FE{F_{\!\!\textnormal{\textsf{eff}}}}
\global\long\def\KE{K_{\textnormal{\textsf{eff}}}}
\global\long\def\E{{\cal E}}
\global\long\def\F{{\cal F}}
\global\long\def\K{{\cal K}}
\global\long\def\EE{{\cal E}_{\textnormal{\textsf{eff}}}}
\global\long\def\fe{f_{\text{\textsf{eff}}}}
\global\long\def\lp{\ell^{\prime}}
\global\long\def\pp{I_{\ul}}
\global\long\def\qq{I_{\lr}}

\DeclareRobustCommand{\ul}{{\raisebox{2pt}{\ytableaushort{ {*(black)} {} , {} {} }}}} 
\DeclareRobustCommand{\ur}{{\raisebox{2pt}{\ytableaushort{ {} {*(black)} , {} {} }}}} 
\DeclareRobustCommand{\ll}{{\raisebox{2pt}{\ytableaushort{ {} {} , {*(black)} {} }}}} 
\DeclareRobustCommand{\lr}{{\raisebox{2pt}{\ytableaushort{ {} {} , {} {*(black)} }}}} 
\DeclareRobustCommand{\di}{{\raisebox{2pt}{\ytableaushort{ {*(black)} {} , {} {*(black)} }}}} 
\DeclareRobustCommand{\of}{{\raisebox{2pt}{\ytableaushort{ {} {*(black)} , {*(black)} {} }}}}
\DeclareRobustCommand{\thr}{{\raisebox{2pt}{\ytableaushort{ {*(black)} {*(black)} , {} {*(black)} }}}}
\DeclareRobustCommand{\thu}{{\raisebox{2pt}{\ytableaushort{ {*(black)} {*(black)} , {*(black)} {} }}}}
\DeclareRobustCommand{\tho}{{\raisebox{2pt}{\ytableaushort{ {} {*(black)} , {*(black)} {*(black)} }}}}
\DeclareRobustCommand{\thd}{{\raisebox{2pt}{\ytableaushort{ {*(black)} {*(black)} , {} {*(black)} }}}}
\DeclareRobustCommand{\rhs}{{\raisebox{2pt}{\ytableaushort{ {} {*(black)} , {} {*(black)} }}}}
\DeclareRobustCommand{\emp}{{\raisebox{2pt}{\ytableaushort{ {} {} , {} {} }}}}
\DeclareRobustCommand{\ub}{{\raisebox{2pt}{\ytableaushort{ {} {*(black)} , {} {*(black)} }}}}
\ytableausetup{boxsize = 2pt}
\newcommand{\ulbig}{\ytableausetup{boxsize = 3pt}
\raisebox{3pt}{\ytableaushort{ {*(black)} {} , {} {} }}
\ytableausetup{boxsize = 2pt}}
\newcommand{\ofbig}{\ytableausetup{boxsize = 3pt}
\raisebox{3pt}{\ytableaushort{ {} {*(black)} , {*(black)} {} }}
\ytableausetup{boxsize = 2pt}}
\newcommand{\urbig}{\ytableausetup{boxsize = 3pt}
\raisebox{3pt}{\ytableaushort{ {} {*(black)} , {} {} }}
\ytableausetup{boxsize = 2pt}}
\newcommand{\llbig}{\ytableausetup{boxsize = 3pt}
\raisebox{3pt}{\ytableaushort{ {} {} , {*(black)} {} }}
\ytableausetup{boxsize = 2pt}}
\newcommand{\lrbig}{\ytableausetup{boxsize = 3pt}
\raisebox{3pt}{\ytableaushort{ {} {} , {} {*(black)} }}
\ytableausetup{boxsize = 2pt}}
\newcommand{\thubig}{\ytableausetup{boxsize = 3pt}
\raisebox{3pt}{\ytableaushort{ {*(black)} {*(black)} , {*(black)} {} }}
\ytableausetup{boxsize = 2pt}}
\newcommand{\empbig}{\ytableausetup{boxsize = 3pt}
\raisebox{3pt}{\ytableaushort{ {} {} , {} {} }}
\ytableausetup{boxsize = 2pt}}
\newcommand{\dibig}{\ytableausetup{boxsize = 3pt}
\raisebox{3pt}{\ytableaushort{ {*(black)} {} , {} {*(black)} }}
\ytableausetup{boxsize = 2pt}}\newcommand{\perttheoryfirst}{Zanardi2014,Zanardi2015,ABFJ}
\newcommand{\perttheorysecond}{Atkins2003,Mirrahimi2008,Reiter2012,Kessler2012a,Zanardi2015a,Azouit2016,azouithal-01394422,Azouit2017,Popkov2018,Das2018,Das2018a,Forni2018}

\title{Dissipative self-interference and robustness of continuous error-correction
to miscalibration}

\author{Victor~V.~Albert}
\email{valbert4@gmail.com}

\selectlanguage{english}%

\affiliation{Institute for Quantum Information and Matter and Walter Burke Institute
for Theoretical Physics, California Institute of Technology, Pasadena,
CA 91125, USA}

\author{Kyungjoo~Noh}

\affiliation{Yale Quantum Institute, Departments of Applied Physics and Physics,
Yale University, New Haven, CT 06520, USA}

\author{Florentin~Reiter}

\affiliation{Department of Physics, Harvard University, Cambridge, MA 02138, USA}
\begin{abstract}
We derive an effective equation of motion within the steady-state
subspace of a large family of Markovian open systems (i.e., Lindbladians)
due to perturbations of their Hamiltonians and system-bath couplings.
Under mild and realistic conditions, competing dissipative processes
destructively interfere without the need for fine-tuning and produce
\textit{no dissipation} within the steady-state subspace. In quantum
error-correction, these effects imply that continuously error-correcting
Lindbladians are robust to calibration errors, including miscalibrations
consisting of operators undetectable by the code. A similar interference
is present in more general systems if one implements a particular
Hamiltonian drive, resulting in a coherent cancellation of dissipation.
On the opposite extreme, we provide a simple implementation of universal
Lindbladian simulation.
\end{abstract}

\date{\today}

\maketitle
Understanding how to reservoir-engineer \cite{Poyatos1996} open quantum
systems is important for the success of noisy intermediate-scale quantum
(NISQ) \cite{Preskill2018} technologies. In this context, one often
encounters the problem of experimentally controlling time-evolution
within a particular subspace of states, e.g., in order to stabilize
states \cite{Altafini2004,Kraus2008,Dirr2009,Ticozzi2009,Krauter2011}
and phases of matter \cite{Diehl2008,Prosen2008a,Verstraete2009,Garcia-Ripoll2009,Dengis2014},
generate gates using Zeno dynamics \cite{Beige2000a,Facchi2002,Arenz2016,S.Touzard},
or protect against unwanted errors \cite{Boulant2005,zoller_stabilizers,Paulisch2015,cats,Leghtas2014,Reiter2017}.
Resolving this problem revolves around variants of either perturbation
theory or adiabatic elimination. In the case of interest here, one
applies a perturbation $\O$ to an unperturbed Lindbladian $\L$ \cite{Belavin1969,Lindblad1976,Gorini1976a,Banks1984,Chruscinski2017}
such that the resulting leading-order time-evolution within the steady-state
subspace of $\L$ is governed by an effective Lindbladian $\LE$.
In general, $\LE$ is difficult to put explicitly in Lindblad form
since there is a complex interplay between dissipation and coherent
evolution inherent in $\L$ \textit{and} arising from $\O$. Cases
in which $\LE$ (to $1^{\text{st}}$ \cite{\perttheoryfirst} or $2^{\text{nd}}$
\cite{\perttheorysecond} order in $\O$) can be simplified are highly
sought after since they yield physical intuition, are numerically
tractable, and provide Hydrogen-atom-like starting points for more
complex scenarios. Due to the aforementioned complexity, such cases
are scarce relative to the many combinations of steady-state structures
\cite{robin}, perturbation types {[}\citealp{baum2}, Sec.~6.1{]},
and features of $\L$ {[}\citealp{thesis}, Sec.~2.1{]}. 

In this Letter, we derive an $\LE$ for arbitrary Hamiltonian and
jump-operator perturbations to certain $\L$ admitting decoherence-free
subspaces (DFS) \cite{Duan1997,Zanardi1997,Lidar1998}, demonstrating
surprising and (to an extent) generic interference effects. Being
an extension of an effective operator formalism (EOF) \cite{Reiter2012}
applicable to a variety of Rydberg \cite{Schempp2015,Schonleber2015,Davis2016},
photonic \cite{Sweke2013,Ramos2016}, and trapped-ion \cite{Reiter2017}
platforms, our formalism and its predicted interference effects should
be observable in and useful to many quantum technologies.

\begin{figure}
\includegraphics[width=1\columnwidth]{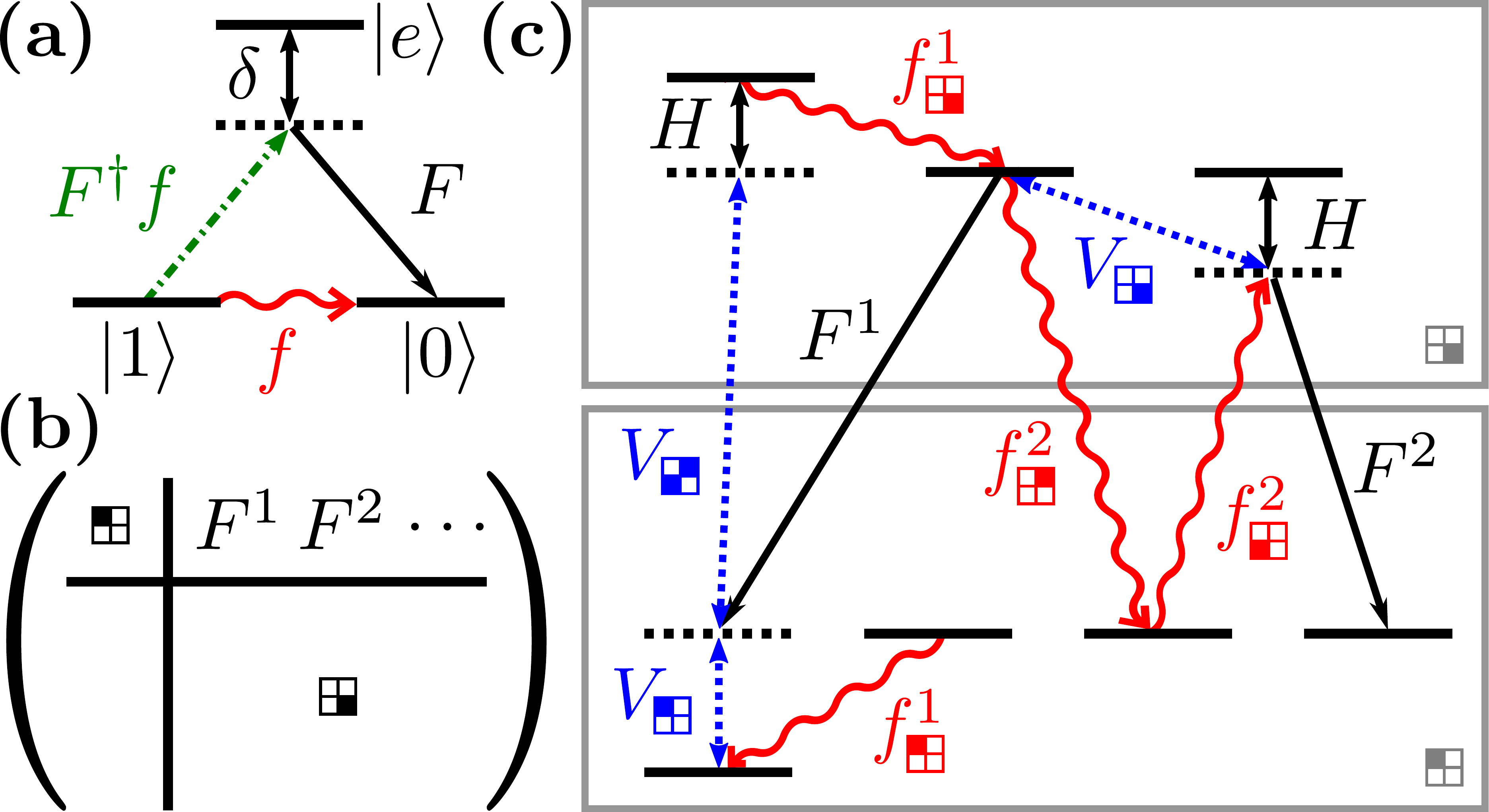}\caption{\label{fig:setup}(color online)\textbf{ (a)} A three-level open system
with unperturbed steady states $\{|0\protect\ket,|1\protect\ket\}$
(forming a DFS), Hamiltonian $H=\protect\d|e\protect\ket\protect\bra e|$,
and jump $F$ (black arrows). To leading order, a jump perturbation
$f$ (red wavy arrow) induces two processes which destructively interfere
with each other {[}see Eq.~(\ref{eq:example}){]}: one is simply
$f$ itself while the other occurs via a virtual transition though
$|e\protect\ket$ via $F^{\protect\dg}f$ (green dotted-dashed arrow).
\textbf{(b)} Sketch of the block matrix formed by jumps $F^{\ell}$
satisfying the condition (\ref{eq:ortho}) necessary for a generalized
interference effect. Each jump operator occupies its own block. The
levels \protect\ulbig represent the DFS while \protect\lrbig are
decaying via the unperturbed Lindbladian $\protect\L=\{H,F^{\ell}\}$
(\ref{eq:setup}). \textbf{(c)} Energy levels of a system satisfying
the assumptions of the EJOF. The perturbations include Hamiltonian
($V$; blue dotted arrows) and jump perturbations $\{f^{1},f^{2}\}$.}
\end{figure}

\textit{Minimal example.---}To gain intuition into the interference
effects, consider first a simple three-level system $\{|0\ket,|1\ket,|e\ket\}$
{[}see Fig.~\ref{fig:setup}(a){]} where the excited level $|e\ket$
resides at an energy $H=\d|e\ket\bra e|$ and decays into $|0\ket$
under jump operator $F=\sqrt{\G}|0\ket\bra e|$ (with corresponding
dissipator $\D[F]\left(\cdot\right)\equiv F\left(\cdot\right)F^{\dg}-\half\{F^{\dg}F,\left(\cdot\right)\}$).
The states $\{|0\ket,|1\ket\}$ form a DFS. Now, assume a small additional
decay $|1\ket\rightarrow|0\ket$ arising from the same coupling to
the bath as the $|e\ket\rightarrow|0\ket$ decay. Under such decay,
$F\rightarrow F+f$ with perturbation $f\equiv\sqrt{\g}|0\ket\bra1|$
and $\G\gg\g$. Naturally, one would think that the leading-order
$O(\g)$ dissipation due to $f$ will be $\D[f]$. However, our formalism
identifies an additional $O(\g)$ effective process that interferes
with this dissipation via the virtual transition $|1\ket\rightarrow|e\ket\rightarrow|0\ket$.
While neither the strong ($F$) nor the weak ($f$) dissipation alone
cause the $|1\ket\rightarrow|e\ket$ part of that transition, perturbing
$F^{\dg}F\rightarrow(F+f)^{\dg}(F+f)$ in $\D[F]$ yields the term
$F^{\dg}f\propto|e\ket\bra1|$ which, when followed by $F$, produces
that transition. That transition is also mediated by the inverse of
the non-Hermitian ``Kamiltonian'' $K=(\d-\frac{i}{2}\G)|e\ket\bra e|$
governing evolution of $|e\ket$. Leading-order dissipative evolution
within the DFS is a superposition of both processes and is governed
by $\LE$ with effective jump operator
\begin{equation}
\FE=f+{\textstyle \frac{i}{2}}FK^{-1}F^{\dg}f=\sqrt{\g}\frac{\d}{\d-\frac{i}{2}\G}|0\ket\bra1|\,.\label{eq:example}
\end{equation}
In the limit of large energy $\d\gg\G$, the virtual $|1\ket\rightarrow|e\ket\rightarrow|0\ket$
transition is off-resonant, the second term in $\FE$ goes to zero,
and one reduces to the intuitive case ($\FE=f$). However, when $\G\gg\d$,
destructive interference between the two terms makes the effective
dissipation disappear entirely ($\FE=0$). Although this cancellation
can be understood nonperturbatively using dark-state physics \footnote{The jump $F+f$ annihilates the dark state $|\protect\psi\protect\rangle=\sqrt{\protect\Gamma}|1\protect\rangle-\sqrt{\protect\gamma}|e\protect\rangle$,
which forms a DFS along with \(|0\protect\ket\). For \(\protect\Gamma\gg\protect\gamma\),
$|\protect\psi\protect\rangle\propto|1\protect\rangle$.}, here we show that the perturbative interference holds much more
generally than previously thought. Generalizing this three-level example,
$\FE=0$ at zero energy for $f=\sqrt{\g}|0\ket\bra\psi|$ with any
$|\psi\ket$. Extending to four or more levels, we will see that $\FE=0$
for a much larger family of $\{F,f\}$.

\textit{Generic cancellation.---}It is uncommon in Hamiltonian perturbation
theory for a correction to be zero for \textit{any} perturbation (unless
a symmetry is present). In this example, we observe such a cancellation
not due to symmetry, but to inherent destructive interference between
generalizations of the two processes discussed above. Consider an
$N+2$-level system $\{|0\ket,|1\ket,|e\ket,|h\ket,\cdots\}$ with
$\{|0\ket,|1\ket\}$ forming a DFS with corresponding projection $\pp=|0\ket\bra0|+|1\ket\bra1|$.
To simplify notation, we partition operators $O$ into four corners
\cite{ABFJ}: $O_{\ul}\equiv\pp O\pp$ acting on the DFS, $O_{\lr}\equiv\qq O\qq$
(with $\qq\equiv1-\pp$) acting on the $N$ decaying states, the ``lowering
operator'' $O_{\ur}\equiv\pp O\qq$ mapping decaying states into
the DFS, and the ``raising operator'' $O_{\ll}\equiv\qq O\pp$ taking
states out of the DFS. Assume no Hamiltonians ($H=0$, for now) and
an unperturbed jump $F=F_{\ur}$, meaning that $F$ maps one directly
into the DFS ($\ulbig$) from the decaying space ($\lrbig$). This
jump can have \textit{any} combination of the $2N$ decay channels
from the $N$ excited states into $\{|0\ket,|1\ket\}$, with the only
restriction that it is surjective,
\begin{equation}
F\left(F^{\dg}F\right)^{-1}F^{\dg}=\pp\,.\label{eq:cond1}
\end{equation}
Interestingly, randomly generated jumps do this: all but a measure-zero
set of $F=F_{\ur}$ consisting of random entries \footnote{See Supplemental Material for a proof of the EJOF and an ancillary
\textsc{Mathematica} file for randomly generated examples.} satisfy (\ref{eq:cond1}). For now, perturb $F$ with any small
$f$ satisfying $f=f_{\thr}^{\ell}$, i.e., any $f$ not mapping $\ulbig$
into $\lrbig$. Applying Eqs.~(\ref{eq:example},\ref{eq:cond1})
yields the effective jump
\begin{align}
\FE & =f_{\ul}-F\left(F^{\dg}F\right)^{-1}F^{\dg}f=f_{\ul}-f_{\ul}=0\label{eq:cancel2}
\end{align}
to leading order in \textit{any} jump perturbation $f_{\thr}$. (We
will later prove that $f_{\rhs}$ doesn't participate at all.) Therefore,
a random jump $F=F_{\ur}$ perturbed by any small perturbation not
mapping out of the DFS generically produces no leading-order dissipation
within the DFS.

This cancellation can be extended to multiple unperturbed jumps $F^{\ell}$,
granted that (\ref{eq:cond1}) holds for each $F^{\ell}$ and the
additional ``orthogonality'' condition
\begin{equation}
F^{\ell}F^{\lp\dg}=\d_{\ell\lp}F^{\ell}F^{\ell\dg}\label{eq:ortho}
\end{equation}
is satisfied. This condition implies that a block matrix consisting
of $\{F^{\ell}\}$ will look like Fig.~\ref{fig:setup}(b). Conditions
(\ref{eq:cond1},\ref{eq:ortho}) imply that $K^{-1}=\sum_{\ell}(-\frac{i}{2}F^{\ell\dg}F^{\ell})^{-1}$
and $F^{\ell}K^{-1}F^{\lp\dg}\propto\pp\d_{\ell\lp}$, yielding once
again no dissipative evolution ($\FE^{\ell}=0$) for \textit{any}
$\{f_{\thr}^{\ell}\}$. Having described the most interesting effect,
we now state our general result---a formalism for tackling perturbations
to a large class of Lindbladians.

\textit{General result.---}Let the unperturbed Lindbladian $\L$
consist of a Hamiltonian $H$ and jump operators $F^{\ell}$,
\begin{equation}
\L\left(\cdot\right)=-i[H,\cdot]+\sum_{\ell}\D[F^{\ell}]\left(\cdot\right)\,.\label{eq:setup}
\end{equation}
Consider coherent and dissipative perturbations, respectively,
\begin{equation}
H\rightarrow H+V\,\,\,\,\,\,\,\,\,\,\,\,\,\,\text{and}\,\,\,\,\,\,\,\,\,\,\,\,\,\,F^{\ell}\rightarrow F^{\ell}+f^{\ell}\,.\label{eq:perts}
\end{equation}
Since $\L$ governs the evolution of a system coupled to a bath \cite{carmichael2,zoller_book,Verstraete2009},
$V$ is a modification of the system Hamiltonian while $f^{\ell}$
modifies the system-bath coupling. If $\L$ is a desired reservoir
engineering operation, then $\{V,f^{\ell}\}$ can be thought of as
uncontrollable coherent evolution and miscalibrations in the engineered
dissipation, respectively. The resulting superoperator perturbation
has terms both $1^{\text{st}}$ and $2^{\text{nd}}$ order in $\{V,f^{\ell}\}$,
$\O=\O_{1}+\O_{2}$, and perturbation theory within the steady-state
subspace yields the Lindbladian {[}\citealp{Macieszczak2015}, Supplement{]}
\begin{equation}
\LE=\P\O\P-\P\O_{1}\L^{-1}\O_{1}\P\,,\label{eq:main}
\end{equation}
where $\L^{-1}$ is the Drazin pseudoinverse {[}\citealp{Zanardi2014},
Eq.~(D4){]} and the asymptotic projection $\P=\I-\L\L^{-1}$ (with
$\I$ identity) projects onto all steady states of $\L$ \cite{ABFJ,thesis}.
The above expression is not particularly illuminating as it is not
in Lindblad form. However, since $\LE$ is a Lindbladian, it must
be expressible in terms of some effective Hamiltonian $\HE$, jump
operators $\FE^{\ell}$, and/or a completely positive (CP) map $\EE$
and its adjoint $\EE^{\ddagger}$ {[}\citealp{Lindblad1976}, Prop.~5{]},
all depending on the unperturbed pieces $\{H,F^{\ell}\}$ and perturbations
$\{V,f^{\ell}\}$. Generally, the expressions may not be simple and
the dependence not explicit, but we are able to express $\L$ in Lindblad
form given the following assumptions. We assume $\L$ admits a unique
DFS $\pp$ and that (A) the unperturbed Hamiltonian acts only on the
decaying subspace ($H=H_{\lr}$) and (B) unperturbed jump operators
map decaying states directly into the DFS ($F^{\ell}=F_{\ur}^{\ell}$).
We assume these hold from now on, noting there are no restrictions
on $\{V,f^{\ell}\}$; see Fig.~\ref{fig:setup}(c) for an example.
To simplify $\LE$, we introduce Kamiltonians
\begin{subequations}
\begin{align}
K & =H-{\textstyle \frac{i}{2}}\sum_{\ell}F^{\ell\dg}F^{\ell}\label{eq:K}\\
\KE & =V_{\of}-{\textstyle \frac{i}{2}}\sum_{\ell}\left(F^{\ell\dg}f_{\ul}^{\ell}+f_{\ul}^{\ell\dg}F^{\ell}\right)\,.\label{eq:Keff}
\end{align}
\end{subequations}
 As we have seen, $K=K_{\lr}$ and its corresponding superoperator
\[
\K(\cdot)\equiv-i\left(K\left(\cdot\right)-\left(\cdot\right)K^{\dg}\right)
\]
govern evolution within $\lrbig$ {[}\citealp{thesis}, Sec.~2.1.3{]}.
As we will see shortly, pieces of the effective Kamiltonian $\KE=(\KE)_{\of}$
map one out of and into the DFS. We picked $\KE$ to depend only on
$\{V_{\of},f_{\ul}\}$ because $\{V_{\ul},f_{\ll}\}$ participate
differently and $\{V_{\lr},f_{\ub}\}$ do not feature to this order.
The resulting simplified $\LE$ (\ref{eq:main}) is as follows \footnotemark[2].
\begin{prop*}
[EJOF]Let $\L$ be a Lindbladian with a unique DFS $\pp$, Hamiltonian
$H_{\lr}$, and jump operators $\{F_{\ur}^{\ell}\}$ (\ref{eq:setup}).
Perturb $\L$ with a Hamiltonian $V$ and jump perturbations $\{f^{\ell}\}$
(\ref{eq:perts}). The effective Lindbladian (\ref{eq:main}) within
the DFS is
\begin{align}
\LE\left(\cdot\right) & =-i[\HE,\left(\cdot\right)]+\sum_{\ell}\D[\FE^{\ell}]\left(\cdot\right)\nonumber \\
 & +\EE\left(\cdot\right)-\half\left\{ \EE^{\ddagger}\left(I\right),\left(\cdot\right)\right\} \,,
\end{align}
where the effective Hamiltonian, jumps, and CP map are
\begin{subequations}
\label{eq:key}
\begin{align}
\HE & =\half\left(V_{\ul}-\KE K^{-1}\KE\right)+H.c.\label{eq:key1}\\
\FE^{\ell} & =f_{\ul}^{\ell}-F^{\ell}K^{-1}\KE\label{eq:key2}\\
\EE\left(\cdot\right) & =-\sum_{\ell,\lp}F^{\lp}\K^{-1}\left(f_{\ll}^{\ell}\left(\cdot\right)f_{\ll}^{\ell\dg}\right)F^{\lp\dg}\,.\label{eq:key3}
\end{align}
\end{subequations}
\end{prop*}
This effective jump-operator formalism (EJOF) reduces to the EOF \cite{Reiter2012}
(see also {[}\citealp{Azouit2016}, Lemma~3{]}) when $f^{\ell}=0$
(and $V_{\lr}=0$). Therefore, the EOF, derived via adiabatic elimination,
can alternatively be derived using time-independent perturbation theory
\footnotemark[2].

The first term (\ref{eq:key1}) represents the resulting coherent
evolution within the DFS. It consists of $V_{\ul}$, a $1^{\text{st}}$-order
effect, and the effective Hamiltonian $\KE K^{-1}\KE+H.c.$ reminiscent
of Hamiltonian perturbation theory. In the latter, $\KE$ (\ref{eq:Keff})
maps states in the kernel ($\ulbig$) of $K$ into the range ($\lrbig$)
using both coherent ($V_{\ll}$) and dissipative ($F^{\dg\ell}f_{\ul}^{\ell}$)
terms, returning via $V_{\ur}$ and $f_{\ul}^{\ell\dg}F$, respectively,
with the ``energy'' denominator determined by $K^{-1}$. Thus there
are cross-terms consisting of leaving via dissipation and returning
via a Hamiltonian and visa-versa. Interestingly, the participating
dissipative perturbation $f_{\ul}^{\ell}$ cannot map one out of the
DFS, instead conspiring with $F^{\dg\ell}$ to provide the dissipative
analogue of $V_{\ll}$. A similar story occurs in the effective jump
$\FE^{\ell}$ (\ref{eq:key2}) and is the key reason behind the highlighted
cancellation. The first part of $\FE^{\ell}$ comes from the first
piece $\P\O_{2}\P=\sum_{\ell}\D[f_{\ul}^{\ell}]$ in Eq.~(\ref{eq:main}),
which is itself a Lindbladian. However, the second piece $-\P\O_{1}\L^{-1}\O_{1}\P$,
which surprisingly is \textit{not} a Lindbladian, contributes the
interference term $F^{\ell}K^{-1}\KE$. This term consists of leaving
the DFS through $(\KE)_{\ll}$ and returning to the DFS via $F^{\ell}$
while paying an ``energy'' penalty determined by the eigenvalues
of $K$. The third term (\ref{eq:key3}) {[}with $\EE^{\ddagger}(I)=\sum_{\ell}f_{\ll}^{\ell\dg}f_{\ll}^{\ell}${]}
results from a nonzero $f_{\ll}^{\ell}$, mapping one out of the DFS
and recovering via $F^{\ell}$, with ``energy'' denominator determined
by the \textit{superoperator} $\K^{-1}\left(\cdot\right)\neq K^{-1}\left(\cdot\right)K^{-1\dg}$.
This term has no analogue in Hamiltonian $2^{\text{nd}}$-order perturbation
theory because it directly connects $\ulbig$ to $\lrbig$ via one
instance of $f_{\ll}$. If $K$ is diagonalizable, we can easily express
$\K^{-1}$ using the eigendecomposition of $K$. However, this formalism
remains valid even for non-diagonalizable $K$ \footnotemark[2].

\textit{Coherent cancellation.---}In our previous examples, we assumed
$H=V=0$ since any initial coherent evolution spoils the interference
effect. We now expand those examples to nonzero Hamiltonians $H\neq0\neq V$,
showing how to restore the interference spoiled by $H$ with a judicious
choice of $V$. We maintain conditions (\ref{eq:cond1},\ref{eq:ortho})
and let $f_{\ll}^{\ell}=0$, so only $\{\HE,\FE^{\ell}\}$ contribute
to $\LE$ (\ref{eq:key}). The presence of $H$ in $K$ (\ref{eq:K})
means that $F^{\ell}K^{-1}F^{\lp\dg}$ is no longer the DFS identity
and $\FE^{\ell}\neq0$. However, since the return to the DFS in $\FE^{\ell}$
occurs via dissipation only, $V_{\ur}$ does not contribute to $\FE$.
Exploiting this effect, we pick
\begin{equation}
V={\textstyle \frac{i}{2}}\sum_{\ell}\left(F^{\ell\dg}f^{\ell}-f^{\ell\dg}F^{\ell}\right)+\tilde{V}\label{eq:cancellationV}
\end{equation}
to cancel the $F^{\ell\dg}f^{\ell}$ term in $\KE$, leaving us with
$(\KE)_{\ll}=\tilde{V}_{\ll}$ that is dependent only on the coherent
perturbation. Picking $\tilde{V}_{\ll}=K\sum_{\ell}(F^{\ell\dg}F^{\ell})^{-1}F^{\ell\dg}f^{\ell}+H.c.$,
the $K$ out front cancels the $K^{-1}$ in $\FE^{\ell}$ and removes
$f_{\ul}^{\ell}$ via the same effect as that in Eq.~(\ref{eq:cancel2}).
In other words, if $H\neq0$, one can use a particular coherent perturbation
to cancel leading-order effects due to unwanted jump perturbations
$f_{\thr}^{\ell}$ \footnotemark[2].

\textit{Universal dissipation.---}In a quick detour from canceling
unwanted dissipation, let us instead use a customizable $V$ to see
what possible dissipation within $\ulbig$ we can generate (c.f. \cite{Zanardi2015a}).
We assume to have full control over the perturbations, showing that
restricting them to $\{V_{\ul},f_{\ul}^{\ell}\}$ allows universal
dissipation within the DFS. First, by letting $\tilde{V}=0$ in Eq.~(\ref{eq:cancellationV}),
we cancel $\KE$-dependent terms in both $\{\HE,\FE^{\ell}\}$ (\ref{eq:key}a-b)
and obtain $\LE=\{V_{\ul},f_{\ul}^{\ell}\}$. Second, letting $d$
be the dimension of the DFS, a general $\LE$ has $d^{2}-1$ jump
operators $\{f^{\ell}\}$. Therefore, if $\L$ has at least $d^{2}-1$
independent jump operators $F^{\ell}$, $\LE$ generates any dissipation
within $\ulbig$.

\textit{Continuous error-correction.}---In conventional QEC, one
starts out in a logical state located in the codespace ($\ulbig$)
and attempts to correct errors caused by an error channel $\E$ by
acting with a recovery channel $\R$. Ideally, $\E$ consists of correctable
noise, so $\R\E=\I$ \cite{nielsen_chuang,preskillnotes}; we focus
on a similar case here. We consider a continuous QEC, where one has
the ability to correct noise via an infinitesimal version of the recovery
$\L=\R-\I$ \cite{Pastawski2011}, whose jumps are the Kraus operators
$\{R^{\ell}\}$ of $\R$ (and we additionally removed the DFS-identity
Kraus operator $R^{0}\propto\pp$). Instead of perturbing $\L$ with
another Lindbladian representing external noise, we consider perturbations
to the jump operators of $\L$, which represent \textit{miscalibration}
of the recovery itself. Such noise is important since a recovery map
is never perfect in real life. It consists of detectable errors $f_{\ll}^{\ell}$
(which we assume are correctable by $\L$), undetectable errors $f_{\ul}^{\ell}$
(which are not correctable since they act nontrivially \textit{within}
the codespace \cite{preskillnotes}), recovery errors $f_{\ur}^{\ell}$,
and correctable errors $f_{\lr}^{\ell}$. The EJOF shows that small
imperfections of \textit{all} types do not harm the quantum information.

Since $\L=\R-\I$ comes from a recovery operation, each jump $F^{\ell}$
is an isometry from a subspace of $\lrbig$ (corresponding to a distinct
error syndrome) into the codespace. Such $F^{\ell}$ automatically
satisfy conditions (\ref{eq:cond1},\ref{eq:ortho}) and, since $\R$
is a channel from $\lrbig$ to $\ulbig$, $\sum_{\ell}F^{\ell\dg}F^{\ell}=\qq$
{[}\citealp{thesis}, Sec.~2.1.4{]}. Let us further assume that miscalibrations
mapping out of the codespace form a channel, $\E\left(\cdot\right)=\sum_{\ell}f_{\ll}^{\ell}\left(\cdot\right)f_{\ll}^{\ell\dg}$,
consisting of correctable noise, i.e., $\R\E(\r)\propto\r$ for all
$\r\in\ulbig$. Application of the EJOF results in the following.
\begin{cor*}
Let $\L=\R-\I$ with corresponding recovery channel $\R\left(\cdot\right)=\sum_{\ell}F^{\ell}\left(\cdot\right)F^{\ell}$
such that $\{F^{\ell}=F_{\ur}^{\ell}\}$ satisfy conditions (\ref{eq:cond1},\ref{eq:ortho}).
Assume small miscalibrations $\{f^{\ell}\}$ in the recovery, $F^{\ell}\rightarrow F^{\ell}+f^{\ell}$,
such that the pieces $\{f_{\ll}^{\ell}\}$ form a noise channel correctable
by $\R$. To leading order, the miscalibrations $\{f^{\ell}\}$ do
not induce errors within the codespace,
\begin{equation}
\LE=0\,.
\end{equation}
\end{cor*}
To prove the above, we have to show that each line in Eq.~(\ref{eq:key})
is zero. First, the simple structure of $\L$ lets us simplify all
Kamiltonian inverses: $K=-{\textstyle \frac{i}{2}}\qq$ and $\K(\r_{\lr})=-\r_{\lr}$
for any $\r_{\lr}$. Plugging this into $\HE$ (\ref{eq:key1}) and
using condition (\ref{eq:ortho}) yields $\HE=0$. Similarly, and
most surprisingly, the interference effect discussed above cancels
the undetectable miscalibrations, yielding $\FE^{\ell}=0$ (\ref{eq:key2}).
Lastly, simplifications to $K$ and the condition on $\{f_{\ll}^{\ell}\}$
yield the trivial CP map (\ref{eq:key3}), $\EE(\r)=-\R\K^{-1}\E(\r)=\R\E(\r)\propto\r$
for all $\r\in\ulbig$.

The above robustness corollary shows that even undetectable miscalibrations
$f_{\ul}^{\ell}$ in continuous error-recovery operations do not affect
the codespace. Qualitatively, it is a statement that holds for any
recovery $\R$ that maps into the codespace after one action and is
applied rapidly (allowing us to consider $\L=\R-\I$). For example,
the statement holds for continuous recoveries for the three-qubit
repetition \cite{Ippoliti2014} and binomial \cite{Jae-MoLihmKyungjooNoh}
codes. For the former, $\pp=|000\rangle\langle000|+|111\rangle\langle111|$,
and its jumps $F^{\ell}=\pp X^{\ell}$ (for qubits $\ell\in\{1,2,3\}$
and $\{X,Y,Z\}$ the usual Pauli matrices) satisfy conditions (\ref{eq:cond1},\ref{eq:ortho}).
Terms $f^{\ell}\propto X^{\ell}$ are corrected by the continuous
recovery while $f^{\ell}\propto Z^{\ell}$ are canceled out due to
interference, despite being undetectable by the code. The terms $f^{\ell}\propto Y^{\ell}$
cannot be corrected since $X^{\ell}$ and $Y^{\ell}$ are not simultaneously
correctable; picking a code correcting both solves this problem.

We cannot make the same statement about all recovery operations since
the assumptions of the EJOF no longer hold. The assumption $F^{\ell}=F_{\ur}^{\ell}$
amounts to the jumps recovering all states (in their range) back into
the codespace \textit{after one action}. Another set of cases is where
$F^{\ell}=F_{\rhs}^{\ell}$ does not map all states immediately into
the codespace, but instead keeps certain states uncorrected (i.e.,
in $\lrbig$) after one action by, e.g., only correcting errors occurring
in a localized region {[}\citealp{thesis}, Sec.~3.4{]}. Such systems
include local recoveries for topological codes and the above corollary
unfortunately \textit{does not apply} to them. Similarly, we cannot
guarantee robustness when there is an inherent Hamiltonian ($H_{\lr}\neq0$).
While we can still use $V$ to coherently cancel any undetectable
miscalibrations $f_{\ul}^{\ell}$ (so that $\FE^{\ell}=0$) as in
the coherent cancellation example above, the presence of $\K^{-1}$
in $\EE$ (\ref{eq:key3}) obstructs us from being able to correct
any detectable errors $f_{\ll}^{\ell}$. So $\LE=0$ only when either
$f_{\ll}^{\ell}=0\neq H$ or visa versa.

\textit{Conclusion.---}We develop an effective jump-operator formalism
to tackle general perturbations to a particular class of Lindbladians
relevant in quantum optics and error correction. We explicitly solve
for the effective Lindbladian $\LE$ governing perturbation-induced
evolution within the steady-state subspace of an unperturbed Lindbladian
$\L$. Using this formalism, we uncover an interference effect that
is a generalized version of the interference observed in dark-state
physics. This interference occurs in generic Lindbladians of the type
we study and can be applied to show that Lindbladian-based error-correction
operations are robust to both detectable \textit{and }undetectable
calibration noise. While this interference is destroyed when the unperturbed
system has a Hamiltonian piece, it can be reinstated with a certain
Hamiltonian perturbation. This formalism also provides a simple way
to realize universal Lindbladian simulation. 

\phfnoteRestoreDefs{origcmds}
\begin{acknowledgments}
We thank Mikhail D. Lukin, Jacob P. Covey, Richard Kueng, John Preskill,
Liang Jiang, Paola Cappellaro, and M\u{a}d\u{a}lin Gu\c{t}\u{a} for
illuminating discussions. We acknowledge financial support from the
Walter Burke Institute for Theoretical Physics at Caltech (V.V.A.),
the Korea Foundation for Advanced Studies (K.N.), and a Feodor-Lynen
fellowship from the Alexander von Humboldt-Foundation (F.R.).
\end{acknowledgments}

\bibliographystyle{apsrev4-1t}
\bibliography{C:/Users/russi/Documents/library}

\clearpage
\onecolumngrid
\appendix
\phfnoteSaveDefs{origcmds}{H,c,k,l,b,d,r,aa,u,v,t,o}

\global\long\def\sqinv{\frac{1}{\sqrt{2}}}
\global\long\def\up{\uparrow}
\global\long\def\do{\downarrow}
\global\long\def\p{\mathcal{P}}
\global\long\def\dqi{\frac{\partial}{\partial q_{i}}}
\global\long\def\dqid{\frac{\partial}{\partial\dot{q}_{i}}}
\global\long\def\a{\alpha}
\global\long\def\b{\beta}
\global\long\def\g{\gamma}
\global\long\def\d{\delta}
\global\long\def\m{\mu}
\global\long\def\n{\nu}
\global\long\def\z{\zeta}
\global\long\def\l{\lambda}
\global\long\def\e{\epsilon}
\global\long\def\x{\chi}
\global\long\def\r{\rho}
\global\long\def\t{\theta}
\global\long\def\c{\csc}
\global\long\def\G{\Gamma}
\global\long\def\D{\mathcal{D}}
\global\long\def\O{\mathcal{O}}
\global\long\def\L{\mathcal{L}}
\global\long\def\T{\mathcal{T}}
\global\long\def\I{1}
\global\long\def\dg{\dagger}
\global\long\def\k{\kappa}
\global\long\def\P{\mathcal{P}_{\!\!\!{\scriptscriptstyle \infty}}}
\global\long\def\R{{\cal R}}
\global\long\def\A{{\cal A}}
\global\long\def\o{\omega}

\global\long\def\LE{{\cal L}_{\textnormal{\textsf{eff}}}}
\global\long\def\HE{H_{\textnormal{\textsf{eff}}}}
\global\long\def\FE{F_{\textnormal{\textsf{eff}}}}
\global\long\def\KE{K_{\textnormal{\textsf{eff}}}}
\global\long\def\KK{{\cal K}_{\textnormal{\textsf{eff}}}}
\global\long\def\E{{\cal E}}
\global\long\def\K{{\cal K}}
\global\long\def\F{{\cal F}}
\global\long\def\V{{\cal V}}
\global\long\def\EE{{\cal E}_{\textnormal{\textsf{eff}}}}
\global\long\def\fe{f_{\text{\textsf{eff}}}}
\global\long\def\lp{\ell^{\prime}}
\global\long\def\pp{I_{\ul}}
\global\long\def\qq{I_{\lr}}

\setcounter{equation}{0}
\renewcommand\theequation{S\arabic{equation}} 

\section*{Appendix: Proof of the EJOF}
\begin{prop*}
[]Let $\L$ be a Lindbladian with a unique DFS $\pp$, Hamiltonian
$H_{\lr}$, and jump operators $\{F_{\ur}^{\ell}\}$. Perturb $\L$
with Hamiltonian $V$ and jump perturbations $\{f^{\ell}\}$. The
effective Lindbladian (\ref{eq:main-1}) within the DFS is 
\begin{align}
\LE\left(\cdot\right) & =-i[\HE,\left(\cdot\right)]+\sum_{\ell}\D[\FE^{\ell}]\left(\cdot\right)+\EE\left(\cdot\right)-\half\left\{ \EE^{\ddagger}\left(I\right),\left(\cdot\right)\right\} \,,\label{eq:leff}
\end{align}
where the effective Hamiltonian, jumps, and CP map are
\begin{subequations}
\label{eq:key-1}
\begin{align}
\HE & =\half\left(V_{\ul}-\KE K^{-1}\KE\right)+H.c.\label{eq:key1-1}\\
\FE^{\ell} & =f_{\ul}^{\ell}-F^{\ell}K^{-1}\KE\label{eq:key2-1}\\
\EE\left(\cdot\right) & =-\sum_{\ell,\lp}F^{\lp}\K^{-1}\left(f_{\ll}^{\ell}\left(\cdot\right)f_{\ll}^{\ell\dg}\right)F^{\lp\dg}\,.\label{eq:key3-1}
\end{align}
\end{subequations}
\end{prop*}
A similar proof of the EOF \cite{Reiter2012} using open-system perturbation
theory was performed in Ref.~\cite{thesis}, Sec.~4.3.5. The adjoint
of a superoperator $\E\left(\cdot\right)=\sum_{i}A_{i}\left(\cdot\right)B_{i}^{\dg}$
is $\E^{\ddagger}\left(\cdot\right)\equiv\sum_{i}A_{i}^{\dg}\left(\cdot\right)B_{i}$.
The perturbation $\O$ to $\L$ consists of contributions from $V$
and $f^{\ell}$ {[}\citealp{baum2}, Sec.~6.1{]}. Let us conveniently
split $\O$ into various superoperators responsible for different
processes. First, define the generalized commutator $\left[A,B\right]^{\star}\equiv AB-BA^{\dg}$
and Kamiltonians $K=H-{\textstyle \frac{i}{2}}\sum_{\ell}F^{\ell\dg}F^{\ell}$
and $\KE\equiv V_{\of}-\frac{i}{2}\sum_{\ell}\left(f_{\ul}^{\ell\dg}F+F^{\ell\dg}f_{\ul}^{\ell}\right)$.
Then, construct the superoperators 
\begin{subequations}
\label{eq:superops}
\begin{align}
\V\left(\cdot\right) & =-i\left[V_{\di}-\frac{i}{2}\sum_{\ell}(f_{\ur}^{\ell\dg}F+F^{\ell\dg}f_{\ur}^{\ell}),\left(\cdot\right)\right]^{\star}\\
\KK\left(\cdot\right) & =-i[\KE,\left(\cdot\right)]^{\star}\\
\F\left(\cdot\right) & =\sum_{\ell}\left(F^{\ell}\left(\cdot\right)f^{\ell\dg}+f^{\ell}\left(\cdot\right)F^{\ell\dg}\right)\,.
\end{align}
\end{subequations}
Split $\O=\O_{1}+\O_{2}$ with $\O_{1}$ containing one instance of
either $f^{\ell}$ or $V$ in each term and $\O_{2}$ containing two:
\begin{subequations}
\label{eq:superops-1}
\begin{align}
\O_{1} & =\V+\KK+\F\\
\O_{2} & =\sum_{\ell}\D[f^{\ell}]\,.
\end{align}
\end{subequations}
Second-order perturbation theory within the DFS yields the effective
Lindbladian {[}\citealp{Macieszczak2015}, Supplement{]}
\begin{equation}
\LE=\P\O\P-\P\O_{1}\K^{-1}\O_{1}\P\equiv\T_{1}+\T_{2}\,.\label{eq:main-1}
\end{equation}
We have simplified $\L^{-1}$ to $\K^{-1}$ in the second term $\T_{2}$
due to the assumption that there is no additional dissipation within
$\lrbig$, $F_{\lr}^{\ell}=0$ {[}\citealp{thesis}, Sec.~2.1.3{]}.
As opposed to Hamiltonian perturbation theory, here the asymptotic
projection $\P$ \cite{ABFJ,thesis} corresponds to a quantum channel
arising from the infinite-time limit of evolution due to $\L$, $\P=\lim_{t\rightarrow\infty}e^{t\L}$.
This channel is trace-preserving, so it is not merely acting on the
DFS since it has to map states initially in $\lrbig$ into the DFS.
We use an analytical formula for it {[}\citealp{ABFJ}, Prop. 3{]},
which for this particular DFS case is
\begin{equation}
\P\left(\cdot\right)=\p_{\ul}\left(\cdot\right)-\p_{\ul}\L\L_{\lr}^{-1}\left(\cdot\right)=\p_{\ul}\left(\cdot\right)-\sum_{\ell}F^{\ell}\K_{\lr}^{-1}\left(\cdot\right)F^{\ell\dg}\,.\label{eq:pinf}
\end{equation}
Above, the four-corners projection superoperators are $\p_{\ul}\left(\cdot\right)=\pp\left(\cdot\right)\pp$
and $\p_{\lr}\left(\cdot\right)=\qq\left(\cdot\right)\qq$, and $\A_{\emp}\equiv\p_{\emp}\A\p_{\emp}$
given any square combination $\empbig$. Above, we have substituted
$\L^{-1}$ for $\K^{-1}$ and used $\p_{\ul}\L\p_{\lr}\left(\cdot\right)=\sum_{\ell}F^{\ell}\left(\cdot\right)F^{\ell\dg}$
{[}\citealp{thesis}, Eq.~(2.8){]}. We use this block notation to
derive the EJOF, introducing the remaining four-corners projectors
$\p_{\ur}\left(\cdot\right)=\pp\left(\cdot\right)\qq$ and $\p_{\ll}\left(\cdot\right)=\qq\left(\cdot\right)\pp$,
noting that they are orthogonal and can add (e.g., $\ofbig\equiv\urbig+\llbig$).
Most importantly, note that 
\begin{equation}
\P=\p_{\ul}\P=\P\p_{\di}=\p_{\ul}\P\p_{\di}\,,\label{eq:asymptotic_projection}
\end{equation}
so $\P$ maps all states into $\ulbig$ and destroys knowledge of
all coherences $\ofbig$ between the DFS and the decaying states.

\subsection{The term $\protect\T_{1}$}

Inserting $\I=\p_{\di}+\p_{\of}$ and using Eq.~(\ref{eq:asymptotic_projection}),
we have
\begin{equation}
\T_{1}=\left(\P\p_{\di}\right)\O\left(\p_{\ul}\P\right)=\P\left(\O_{\ul}+\p_{\lr}\O\p_{\ul}\right)\P=\O_{\ul}+\P\p_{\lr}\O\p_{\ul}\,,
\end{equation}
so we only need two superoperator elements, $\O_{\ul}$ and $\p_{\lr}\O\p_{\ul}$,
for this term. Note that we have applied $\P\p_{\ul}=\p_{\ul}$ and
replaced the rightmost $\P$ with $\p_{\ul}$ since the states we
are perturbing are in $\ulbig$. The former element is a projection
of $\O$ onto the DFS while the latter is a leakage term into the
decaying space. These elements are listed below for all of the terms
$\A\in\{\V,\KK,\F,\D[f^{\ell}]\}$ of $\O$. 
\begin{center}
\begin{tabular}{ccc}
\toprule 
$\A$ & $\A_{\ul}$ & $\p_{\lr}\A\p_{\ul}$\tabularnewline
\midrule
$\V$ & $-i[V_{\ul},\left(\cdot\right)]$ & 0\tabularnewline
$\KK$ & 0 & 0\tabularnewline
$\F$ & 0 & 0\tabularnewline
$\D[f^{\ell}]$ & ~~~~~~~~~~$\D[f_{\ul}^{\ell}]\left(\cdot\right)-\half\{f_{\ll}^{\ell\dg}f_{\ll}^{\ell},\left(\cdot\right)\}$~~~~~~~~~~ & $f_{\ll}^{\ell}\left(\cdot\right)f_{\ll}^{\ell\dg}$\tabularnewline
\bottomrule
\end{tabular}
\par\end{center}

Luckily, $\p_{\lr}\A\p_{\ul}=0$ for $\A\in\{\K,\V\}$ due to the
fact that their constituents act from one side at a time (the no-leak
property; see {[}\citealp{ABFJ}, Sec.~I.B{]}). Also, $\p_{\lr}\F\p_{\ul}=0$
since its constituent $F^{\ell}=F_{\ur}^{\ell}$ cannot map one into
$\lrbig$ by construction. From the above table, we see that $\V$
contributes the first term in $\HE$ (\ref{eq:key1-1}) and $\D[f^{\ell}]$
contributes the dissipator $\D[f_{\ul}^{\ell}]$. We cannot yet combine
all $f_{\ll}^{\ell}$ terms because we still need to act on $\p_{\lr}\D[f^{\ell}]\p_{\ul}$
with $\P$ (\ref{eq:pinf}):
\begin{equation}
\sum_{\ell}\P\p_{\lr}\D[f^{\ell}]\p_{\ul}\left(\cdot\right)=-\sum_{\ell,\lp}F^{\lp}\K_{\lr}^{-1}\left(f_{\ll}^{\ell}\left(\cdot\right)f_{\ll}^{\ell\dg}\right)F^{\lp\dg}\equiv\EE\left(\cdot\right)\,.\label{eq:Eeff}
\end{equation}
This provides the first term for the $\EE$-dependent part of $\LE$
(\ref{eq:leff}). To complete the derivation of $\T_{1}$, we need
to prove that $\EE^{\ddagger}(I)=\sum_{\ell}f_{\ll}^{\ell\dg}f_{\ll}^{\ell}$.
The anticommutator term should be $\EE^{\ddagger}(\pp)$ since $\EE$
is a channel from $\ulbig$ to itself, but padding with $\qq$ doesn't
make any difference and looks simpler. Note that $H=H_{\lr}$ commutes
with $\qq$ and so $\K^{\ddagger}(\qq)=-\sum_{\lp}F^{\lp\dg}F^{\lp}$.
Plugging this into $\EE^{\ddagger}(I)$ cancels the $\K^{-1\ddagger}$,
yielding
\begin{align}
\EE^{\ddagger}(I) & =\sum_{\ell}f_{\ll}^{\ell\dg}\K^{-1\ddagger}\left(-\sum_{\lp}F^{\lp\dg}F^{\lp}\right)f_{\ll}^{\ell}=\sum_{\ell}f_{\ll}^{\ell\dg}\K^{-1\ddagger}\K^{\ddagger}(\qq)f_{\ll}^{\ell}=\sum_{\ell}f_{\ll}^{\ell\dg}f_{\ll}^{\ell}\,.\label{eq:Eeff1}
\end{align}
This provides the anticommutator term for the $\EE$-dependent part
of $\LE$ (\ref{eq:leff}). We are left with the $\KE$-dependent
terms in $\HE$ (\ref{eq:key1-1}) and $\FE$ (\ref{eq:key2-1}),
which come from $\T_{2}$.

\subsection{The term $\protect\T_{2}$}

This term is more difficult since two actions of the perturbation
are present. We likewise need to determine which superoperator elements
are required for the calculation. Since $\K^{-1}$ does not act on
$\ulbig$ ($\K^{-1}=\K_{\tho}^{-1}$), the first part of $\T_{2}$
is $\K^{-1}\O_{1}\P=\K^{-1}\left(\p_{\of}\O_{1}\p_{\ul}+\p_{\lr}\O_{1}\p_{\ul}\right)\P$.
However, we can see that $\p_{\lr}\O_{1}\p_{\ul}=0$ from the previous
table, so only $\K_{\of}^{-1}$ participates. Inserting this into
$\T_{2}$ and using Eq.~(\ref{eq:asymptotic_projection}) yields
\begin{equation}
\T_{2}=-\left(\p_{\ul}\O_{1}\p_{\of}+\P\p_{\lr}\O_{1}\p_{\of}\right)\K^{-1}\left(\p_{\of}\O_{1}\p_{\ul}\right)\,.\label{eq:t2}
\end{equation}
Therefore, three elements are relevant; they are listed in the table
below for all of the terms $\A\in\{\V,\KK,\F\}$ of $\O_{1}$:
\begin{center}
\begin{tabular}{cccc}
\toprule 
$\A$ & $\p_{\of}\A\p_{\ul}$ & $\p_{\lr}\A\p_{\of}$ & $\p_{\ul}\A\p_{\of}$\tabularnewline
\midrule
$\V$ & 0 & 0 & 0\tabularnewline
$\KK$ & ~~~~$-i[(\KE)_{\ll},\left(\cdot\right)]^{\star}$~~~~ & $-i[(\KE)_{\ll},\left(\cdot\right)]^{\star}$ & $-i[(\KE)_{\ur},\left(\cdot\right)]^{\star}$\tabularnewline
$\F$ & 0 & 0 & ~~~~$\sum_{\ell}\left(F^{\ell}\left(\cdot\right)f_{\ul}^{\ell\dg}+f_{\ul}^{\ell}\left(\cdot\right)F^{\ell\dg}\right)$\tabularnewline
\bottomrule
\end{tabular}
\par\end{center}

The first part $\K^{-1}\left(\p_{\of}\O_{1}\p_{\ul}\right)$ in $\T_{2}$
(\ref{eq:t2}) is shared by all terms, so we simplify it first by
noting that the superoperator inverse $\K_{\of}^{-1}$ can be written
in terms of operator inverses due to the restriction $F^{\ell}=F_{\ur}^{\ell}$
{[}\citealp{thesis}, Eq.~(2.8){]},
\begin{equation}
\K_{\of}^{-1}\left(\cdot\right)=\K_{\ur}^{-1}\left(\cdot\right)+\K_{\ll}^{-1}\left(\cdot\right)=-i\left(\cdot\right)K^{-1\dg}+iK^{-1}\left(\cdot\right)=i[K^{-1},\left(\cdot\right)]^{\star}\,.
\end{equation}
Plugging this and the first column of the above table into the first
part of $\T_{2}$ yields
\begin{equation}
\K^{-1}\left(\p_{\of}\O_{1}\p_{\ul}\right)\left(\cdot\right)=\left[K^{-1},[(\KE)_{\ll},\left(\cdot\right)]^{\star}\right]^{\star}=K^{-1}\KE\left(\cdot\right)+H.c.\,,
\end{equation}
where we remember that the state $\left(\cdot\right)\in\ulbig$ and
only $(\KE)_{\ll}$ can map $\left(\cdot\right)$ into $\lrbig$ (so
that $K^{-1}$ acts on the result). In the last equality, we let $(\KE)_{\ll}\rightarrow\KE$
since adding $(\KE)_{\ur}$ does not make any difference, i.e., $K^{-1}(\KE)_{\ur}=0$.
Now let us plug this simplified first part as well as all of the nonzero
terms from the table into $\T_{2}$:
\begin{equation}
\T_{2}=-\left(\p_{\ul}\KK\p_{\of}+\p_{\ul}\F\p_{\of}+\P\p_{\lr}\KK\p_{\of}\right)\left(K^{-1}\KE\left(\cdot\right)+H.c.\right)\,.
\end{equation}
We now determine the contribution coming from each of the three terms
in the leftmost parentheses. Using the above table and substituting
$(\KE)_{\ur}\rightarrow\KE$ in the first line below, the first two
terms are simple:
\begin{subequations}
\label{eq:superops-1-1}
\begin{align}
-\p_{\ul}\KK\p_{\of}\left(K^{-1}\KE\left(\cdot\right)+H.c.\right) & =i\left[\KE K^{-1}\KE,\left(\cdot\right)\right]^{\star}\\
-\p_{\ul}\F\p_{\of}\left(K^{-1}\KE\left(\cdot\right)+H.c.\right) & =-\sum_{\ell}\left(F^{\ell}K^{-1}\KE\left(\cdot\right)f_{\ul}^{\ell\dg}+H.c.\right)\,.
\end{align}
\end{subequations}
For the third term, note first this curious formula that we will use
to eliminate the inverse coming from $\P\p_{\lr}$: 
\begin{equation}
\K^{-1}\left(i[K^{-1},\left(\cdot\right)]^{\star}\right)=\K^{-1}\K\left(K^{-1}\left(\cdot\right)K^{-1\dg}\right)=K^{-1}\left(\cdot\right)K^{-1\dg}
\end{equation}
for any operator $\left(\cdot\right)\in\lrbig$. Plugging in Eq.~(\ref{eq:pinf})
and applying the above formula yields
\begin{equation}
-\P\p_{\lr}\KK\p_{\of}\left(K^{-1}\KE\left(\cdot\right)+H.c.\right)=\sum_{\ell}F^{\ell}K^{-1}\KE\left(\cdot\right)\KE^{\dg}K^{-1\dg}F^{\ell\dg}\,.\label{eq:both-sides}
\end{equation}

\subsection{Combining $\protect\T_{1}$ and $\protect\T_{2}$}

Plugging the $\EE$-dependent terms (\ref{eq:Eeff},\ref{eq:Eeff1}),
all $V_{\ul}$- and $f_{\ul}^{\ell}$-dependent terms in the first
table above, and Eqs.~(\ref{eq:superops-1-1},\ref{eq:both-sides})
yields the effective Lindbladian
\begin{align}
\LE\left(\cdot\right) & =-i[\HE,\left(\cdot\right)]+\EE\left(\cdot\right)-\left\{ \EE^{\ddagger}(\pp),\left(\cdot\right)\right\} +{\textstyle \frac{i}{2}}\left\{ \KE K^{-1}\KE-H.c.,\left(\cdot\right)\right\} \nonumber \\
 & +\sum_{\ell}\D[f_{\ul}^{\ell}]\left(\cdot\right)+F^{\ell}K^{-1}\KE\left(\cdot\right)\KE^{\dg}K^{-1\dg}F^{\ell\dg}-\left(F^{\ell}K^{-1}\KE\left(\cdot\right)f_{\ul}^{\ell\dg}+H.c.\right)\,,
\end{align}
where we have absorbed the Hermitian part of $\KE K^{-1}\KE$ into
$\HE$. Remarkably, the last term in the second line and the third
line simplify to $\sum_{\ell}\D[\FE^{\ell}]$. Collecting all of the
terms in the second line that act nontrivially from both sides into
$\FE$ makes this more clear, leaving only the anticommutator term
$\sum_{\ell}\FE^{\ell\dg}\FE^{\ell}$ to be determined from the last
term below:
\begin{align}
\!\!\!\!\LE\left(\cdot\right) & =-i[\HE,\left(\cdot\right)]+\EE\left(\cdot\right)-\{\EE^{\ddagger}(\pp),\left(\cdot\right)\}+\sum_{\ell}\FE^{\ell}\left(\cdot\right)\FE^{\ell\dg}-{\textstyle \frac{1}{2}}\left\{ \sum_{\ell}f_{\ul}^{\ell\dg}f_{\ul}^{\ell}-i\left(\KE K^{-1}\KE-H.c.\right),\left(\cdot\right)\right\} \,.
\end{align}
Let us now write $\KE=V_{\ll}+\frac{i}{2}G_{\ll}+H.c.$, where $G_{\ll}=\sum_{\ell}F^{\ell\dg}f_{\ul}^{\ell}$.
We abbreviate $V_{\ll}^{\dg}\equiv(V_{\ll})^{\dg}$ and similarly
for $G$. Plugging this into $\KE$ and simplifying yields
\begin{align}
-i\left(\KE K^{-1}\KE-H.c.\right)= & -iV_{\ll}^{\dg}\left(K^{-1}-K^{-1\dg}\right)V_{\ll}+{\textstyle \frac{i}{4}}G_{\ll}^{\dg}\left(K^{-1}-K^{-1\dg}\right)G_{\ll}\nonumber \\
 & -\half G_{\ll}^{\dg}\left(K^{-1}+K^{-1\dg}\right)V_{\ll}-\half V_{\ll}^{\dg}\left(K^{-1}+K^{-1\dg}\right)G_{\ll}\,.
\end{align}
We obtain the same for $\sum_{\ell}\FE^{\ell\dg}\FE^{\ell}-f_{\ul}^{\ell\dg}f_{\ul}^{\ell}$
to finish the proof. For this, we have to use another curious identity
that is proven using the definition of $K$,
\begin{equation}
\sum_{\ell}K^{-1\dg}F^{\ell\dg}F^{\ell}K^{-1}=-i\left(K^{-1}-K^{-1\dg}\right)\,.
\end{equation}
Plugging this in, splitting $\KE$ into $V_{\ll}$ and $G_{\ll}$,
and simplifying yields
\begin{equation}
\sum_{\ell}\FE^{\ell\dg}\FE^{\ell}-f_{\ul}^{\ell\dg}f_{\ul}^{\ell}=-i\left(\KE K^{-1}\KE-H.c.\right)\,.
\end{equation}

\end{document}